\documentclass[aps,prl,superscriptaddress,showpacs,twocolumn,
]{revtex4-1}
\usepackage{graphicx}
\usepackage{epstopdf}
\usepackage{natbib}

\begin{document}


\title{Lifetimes of antiferromagnetic magnons in two and three dimensions: experiment, theory, and numerics}


\author{S.P. Bayrakci}
\email[]{bayrakci@fkf.mpg.de}
\affiliation{Max-Planck-Institut f\"ur Festk\"orperforschung, Heisenbergstrasse 1, D-70569 Stuttgart, Germany}
\author{D.A. Tennant}
\affiliation{Helmholtz-Zentrum Berlin f\"ur Materialien und Energie, Hahn-Meitner-Platz 1, D-14109 Berlin, Germany}
\affiliation{Institut f\"ur Festk\"orperphysik, Technische Universit\"at Berlin, Hardenbergstrasse 36, D-10623 Berlin, Germany}
\author{Ph. Leininger}
\affiliation{Max-Planck-Institut f\"ur Festk\"orperforschung, Heisenbergstrasse 1, D-70569 Stuttgart, Germany}
\author{T. Keller}
\affiliation{Max-Planck-Institut f\"ur Festk\"orperforschung, Heisenbergstrasse 1, D-70569 Stuttgart, Germany}
\author{M.C.R. Gibson}
\affiliation{Helmholtz-Zentrum Berlin f\"ur Materialien und Energie, Hahn-Meitner-Platz 1, D-14109 Berlin, Germany}
\author{S.D. Wilson}
\affiliation{Materials Science Division, Lawrence Berkeley National Laboratory, Berkeley, CA 94720, USA}
\author{R.J. Birgeneau}
\affiliation{Materials Science Division, Lawrence Berkeley National Laboratory, Berkeley, CA 94720, USA}
\author{B. Keimer}
\affiliation{Max-Planck-Institut f\"ur Festk\"orperforschung, Heisenbergstrasse 1, D-70569 Stuttgart, Germany}


\date{\today}

\begin{abstract}
A high-resolution neutron spectroscopic technique is used to measure momentum-resolved magnon lifetimes in the prototypical two- and three-dimensional antiferromagnets Rb$_{2}$MnF$_{4}$ and MnF$_{2}$, over the full Brillouin zone and a wide range of temperatures. We rederived theories of the lifetime resulting from magnon-magnon scattering, thereby broadening their applicability beyond asymptotically small regions of wavevector and temperature. Corresponding computations, combined with a small contribution reflecting collisions with domain boundaries, yield excellent quantitative agreement with the data.
\end{abstract}

\pacs{75.30.Ds, 75.40.Gb, 75.50.Ee}

\maketitle


Dynamics in Heisenberg antiferromagnets have been studied very extensively by means of theoretical approaches such as hydrodynamics\cite{halperin1}, dynamical scaling\cite{halperin2}, and interacting spin-wave methods\cite{dyson}.  In the dilute magnon gas in two and three dimensions \cite{harris,tyc,cottam,bohnen,kopietz,balcar}, the dominant damping processes are identified by diagrammatic perturbation theory to be elastic pair-wise collisions of magnons (see Fig. 1a-b); the transition rate is given by Fermi's Golden Rule.  Analytic formulae for the relaxation as a function of wavevector $\mathit{q}$ and temperature $\mathit{T}$ were derived for this mechanism (see Fig. 1).  Following the discovery of high-temperature superconductivity in doped two-dimensional (2D) antiferromagnets, recent theoretical work has focused specifically on the differences between the behavior of magnons in two and three dimensions\cite{tyc,tyc2,kopietz}.  Surprisingly, in recent experimental studies of both 2D \cite{huberman} and 3D \cite{bayrakci} antiferromagnets, the measured linewidths disagreed considerably with predicted values, and anomalous additional line broadening at low temperatures that eluded explanation was observed.  To address these issues, we have reexamined this decades-old problem both experimentally and theoretically, in both 2D and 3D. 

\begin{figure}
\includegraphics[keepaspectratio=true,clip=true,trim=51mm 57mm 47mm 88mm,width=86mm]{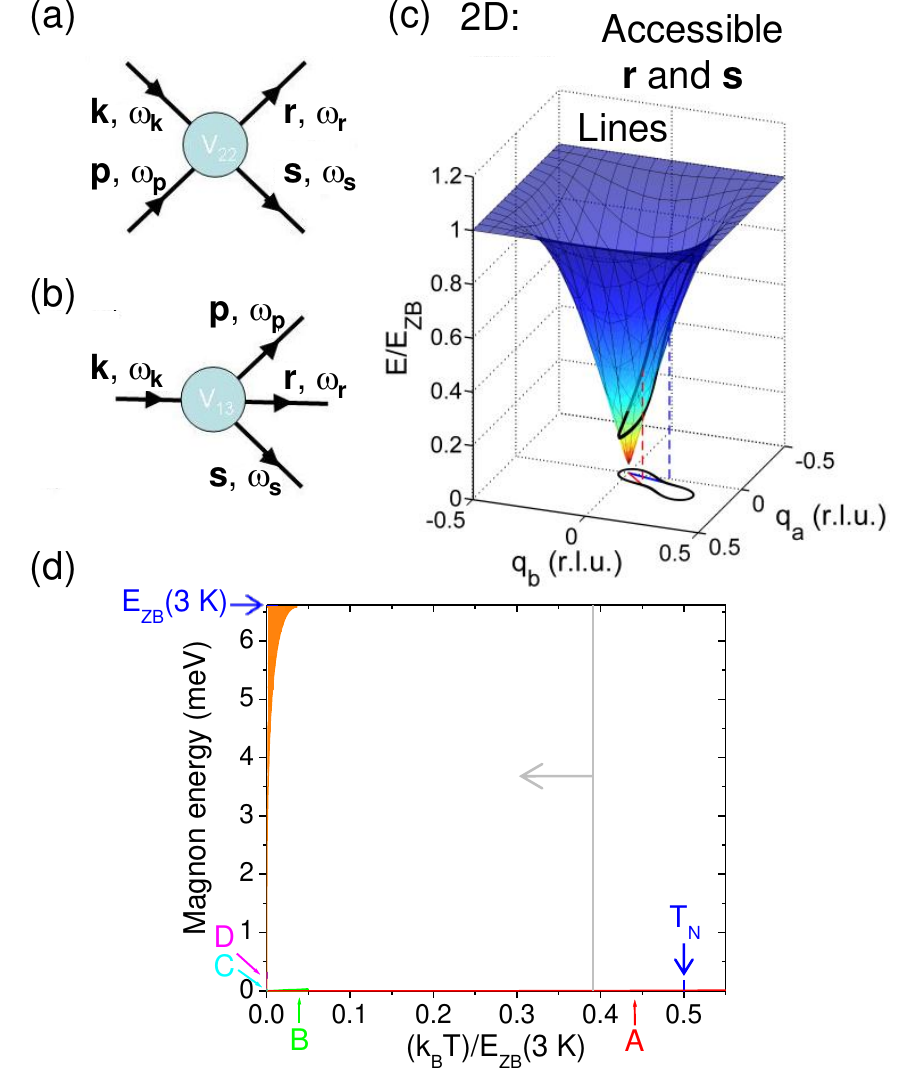}  
\caption{\label{1} (Color online) (a),(b) The primary scattering processes involve four magnons.  (a) The two-in/two-out process dominates at low temperatures. (b) The one-in/three-out (or three-in/one-out) process is calculated to be negligible compared to that of (a).  (c) The dispersion surface in 2D.  The lifetime broadening is determined by the number of accessible final states.  These scattered states lie on loops on the surface, shown here for $\mathbf{k} = (0.1,0.1)$ and $\mathbf{p} = (0.05, 0.2)$.  The loop is also projected onto the Brillouin zone.  Strict conservation of energy and momentum results in the loop being a delta function; lifetime broadening relaxes this condition.  For small (large) $\mathit{k}$, the loop shrinks (grows), resulting in narrow (few scattering states available) and broad (many available) linewidths, approximately as $\Gamma_{\mathrm{2D}} \sim \mathit{k}$.  In 3D, the scattered states are in the surfaces, or ``shells'', of 3D volumes, with $\Gamma_{\mathrm{3D}} \sim \mathit{k}^{2}$.  (d) For illustrative purposes, we have concretized the definition of the four regimes for which TH provide analytical solutions in 2D by stipulating that a strict inequality represents a factor of 10 difference in magnitude.  The corresponding boundaries of the regimes A-D \cite{tyc} are indicated.  The grey line indicates the highest temperature at which we measured linewidths.  The region of applicability of Kopietz' analytical result \cite{kopietz} is shown in orange; the upper limit of the temperature does not quite overlap with the lowest temperature at which we measured (3 K).  Analogous regimes are employed by HKHH in 3D; their area is comparable to that shown for TH.}
\end{figure}

Rb$_{2}$MnF$_{4}$\cite{cowley} and MnF$_{2}$\cite{erickson} are near-ideal realizations of 2D and 3D antiferromagnets, respectively.  Due to their structural and chemical similarity, they are a well-suited pair of compounds for comparison of linewidths in 2D and 3D.  In these body-centered tetragonal systems, the magnetic interactions are dominated by the antiferromagnetic (AF) exchange coupling  $\mathit{J}$  between $\mathit{S} = 5/2$ Mn$^{2+}$ spins: for Rb$_{2}$MnF$_{4}$, between the 4 nearest neighbors in the $\mathit{ab}$-plane, and for MnF$_{2}$, between the 8 next-nearest neighbors (arranged at the corners of a tetragonal prism).  A uniaxial anisotropy parameter $\mathit{D}$ arising primarily from dipole-dipole interactions causes the spins in both compounds to align along the $\mathit{c}$-axis.  The resulting dispersion relations in the two systems are very similar: the zone-boundary energy and energy gap at 4 K are 6.6 meV (along $(\mathit{qq}0)$) and 0.6 meV in Rb$_{2}$MnF$_{4}$, respectively\cite{cowley}, and 6.3 meV (along $(\mathit{q}00)$) and 1.1 meV for MnF$_{2}$\cite{low}.  The N\'{e}el temperature $\mathit{T}_{\mathit{N}}$ is 38.4 K in Rb$_{2}$MnF$_{4}$ and 67.6 K in MnF$_{2}$.  

We used the high-resolution NRSE-TAS (neutron resonance spin-echo triple-axis spectroscopy) technique\cite{kellerNRSE} to measure magnon linewidths in the $\mathit{ab}$-plane of these antiferromagnets (see Figs. 2-4).  Temperatures ranged from 3 K up to 0.8 $\mathit{T}_{\mathit{N}}$ (Rb$_{2}$MnF$_{4}$) and 0.6 $\mathit{T}_{\mathit{N}}$ (MnF$_{2}$); the wavevectors spanned the full magnetic Brillouin zone (BZ).  The neutron spectrometer and experimental methods used are as described in Ref. 12.  The mosaicity of the samples used was measured by means of gamma-ray diffractometry and found to be 0.44' and 0.99' for MnF$_{2}$ and Rb$_{2}$MnF$_{4}$, respectively.  The MnF$_{2}$ sample was oriented in the (HK0) scattering plane and the Rb$_{2}$MnF$_{4}$ sample in the (HHL) scattering plane.  The magnetic BZ used in MnF$_{2}$ was that centered around (010) and that in Rb$_{2}$MnF$_{4}$ around ($\frac{1}{2}$$\frac{1}{2}$0) (the latter in tetragonal crystal coordinates with $\mathit{a} = 4.216 \mathrm{\AA}$; this reflects K$_{2}$NiF$_{4}$-type magnetic ordering\cite{cowley}).  The final neutron wavevectors $\mathit{k}_{\mathit{f}}$ used ranged from 1.7 to 3.2 $\mathrm{\AA}^{-1}$.  

In NRSE-TAS magnon linewidth measurements, accurate knowledge of the magnon dispersion is required in order to set the angles of the RF tilt coils correctly\cite{keller,bayrakci}.  Correspondingly, we measured the dispersion at all temperatures at which we measured linewidths (except where the change with temperature was not experimentally resolvable).

The results for both compounds are qualitatively similar to those found for $\mathit{q}_{\mathit{c}}$-magnons in MnF$_{2}$\cite{bayrakci}.  A broad peak at 3 K of maximum value $\sim$6 $\mu$eV at small $\mathit{q}$ characterizes both data sets.  We initially compared the results with analytical approximations for the magnon linewidth originating from two-in/two-out magnon collisions (see Fig. 1a).  The corresponding expressions derived by Harris et al. (HKHH) \cite{harris} and Tyc and Halperin (TH) \cite{tyc} apply to restricted regions of small wavevectors at low temperatures (see Fig. 1d).  Their general low-temperature expressions, which contain no adjustable parameters and involve summations over the BZ, are intended for use at long wavelengths. Umklapp processes, which become increasingly important as $\mathit{q}$ and $\mathit{T}$ increase, are not treated properly in the scattering matrix elements.  

We rederived the scattering matrix elements for general $\mathit{q}$, treating Umklapp processes correctly\cite{HKHHnote}, and then computed the corresponding linewidths $\Gamma_{\mathrm{mag}}(\mathit{q},\mathit{T})$ numerically.  It became clear that ``on-shell'' computations do not give correct line broadenings, particularly near the zone boundaries, because of the curvature of the dispersion relation and the large density of states present there.  For this reason, we evaluated $\Gamma_{\mathrm{mag}}(\mathit{q},\mathit{T})$ self-consistently.  We incorporated experimentally-measured values of the renormalized magnon energies into the calculation, and hence our results can be applied at higher temperatures than those considered by HKHH and TH.  Through such calculations, we determined that incorrect treatment of Umklapp processes yields inaccurate linewidth results, even at small wavevectors and low temperatures.  Independent of how these are treated, our numerical results differ considerably in magnitude from the analytical approximations, and thus numerical evaluations are indispensable even for the limiting cases of small $\mathit{q}$ and low $\mathit{T}$.   

The computations involve sums over $\mathbf{p}$ and $\mathbf{q}$ states, where $\mathbf{k} + \mathbf{p} = \mathbf{r} + \mathbf{s}$, $\mathbf{r} = \mathbf{p} - \mathbf{q}$, and $\mathbf{s} = \mathbf{k} + \mathbf{q}$.  For $\mathbf{p}$, a grid with density of $(81/(2\pi))^{2} \mathit{pts}\cdot \mathit{a}^{2}$, with 64 times higher density in the central 1/8$^{\mathrm{th}}$ of the BZ, is used in the summation, weighted by the reciprocal space volume per point.  The grid in $\mathbf{q}$ has a uniform density of  $(81/(2\pi))^{2} \mathit{pts}\cdot \mathit{a}^{2}$.  Different grid sizes were checked to ensure that convergence occurred.  Momentum conservation in the scattering process is achieved by the choices $\mathbf{r} = \mathbf{p} - \mathbf{q}$ and $\mathbf{s} = \mathbf{k} + \mathbf{q}$.  To approximate the delta-function energy-acceptance window, a Lorentzian probability distribution $p(\Delta)=1/(\pi\gamma[1+(\Delta/\gamma)^{2}])$ is used, where $\Delta = \omega(\mathit{k}) +\omega(\mathit{p}) - \omega(\mathit{r}) - \omega(\mathit{s})$ is the deviation from the conserved result, and the width $\gamma$ is determined by the lifetimes of the ``intermediate'' spin-waves involved in the scattering process, $\mathit{viz.}$ $\gamma = \gamma_{p} +\gamma_{r} + \gamma_{s}$ \cite{harris}, where the intermediate line broadenings are interpolated for the relevant wavevectors from those previously computed.  To speed up the calculation, the Lorentzian is truncated at $\pm 4\gamma$, with its amplitude adjusted correspondingly to preserve unit area.  Self-consistent calculations are performed by choosing an initial narrow linewidth that is uniform in wavevector and then iterating.  Iterations were performed until the results converged to within 0.4\% (for MnF$_{2}$) or 1\% (for Rb$_{2}$MnF$_{4}$), over the full range of wavevector and temperature.

\begin{figure}
\includegraphics[keepaspectratio=true,clip=true,trim=48mm 60mm 55mm 46mm,width=86mm]{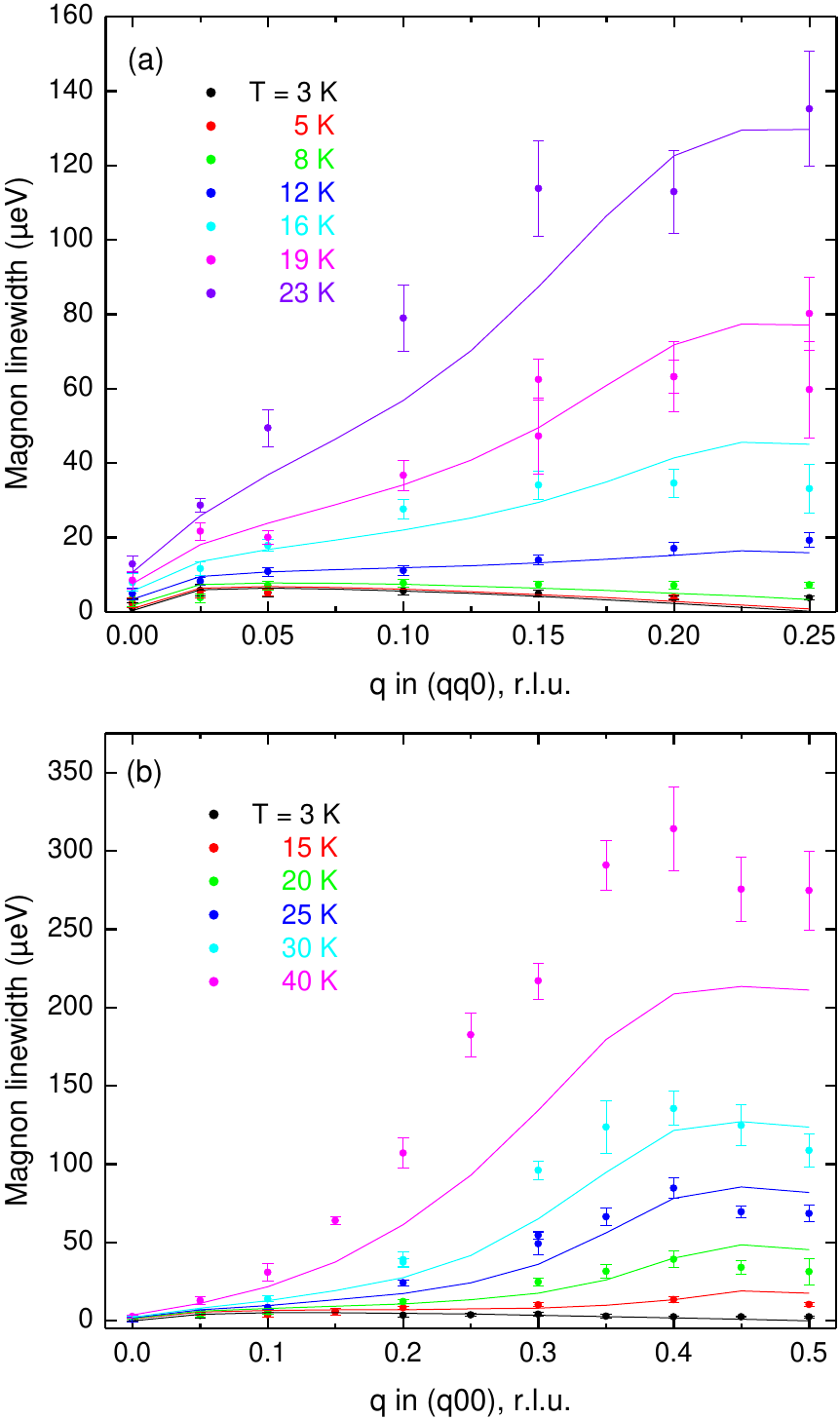}
\caption{\label{2} (Color online) Linewidth data over the full BZ in (a) Rb$_{2}$MnF$_{4}$ and (b) MnF$_{2}$.  The curves show self-consistent calculations, with no free parameters, of 4-magnon scattering, added to a small, nearly $\mathit{T}$-independent contribution, peaked at low $\mathit{q}$, representing 2-magnon scattering from crystal blocks (with one free parameter, fitted to the 3 K data).}
\end{figure}

\begin{figure}
\includegraphics[keepaspectratio=true,clip=true,trim=50mm 102mm 54mm 78mm,width=86mm]{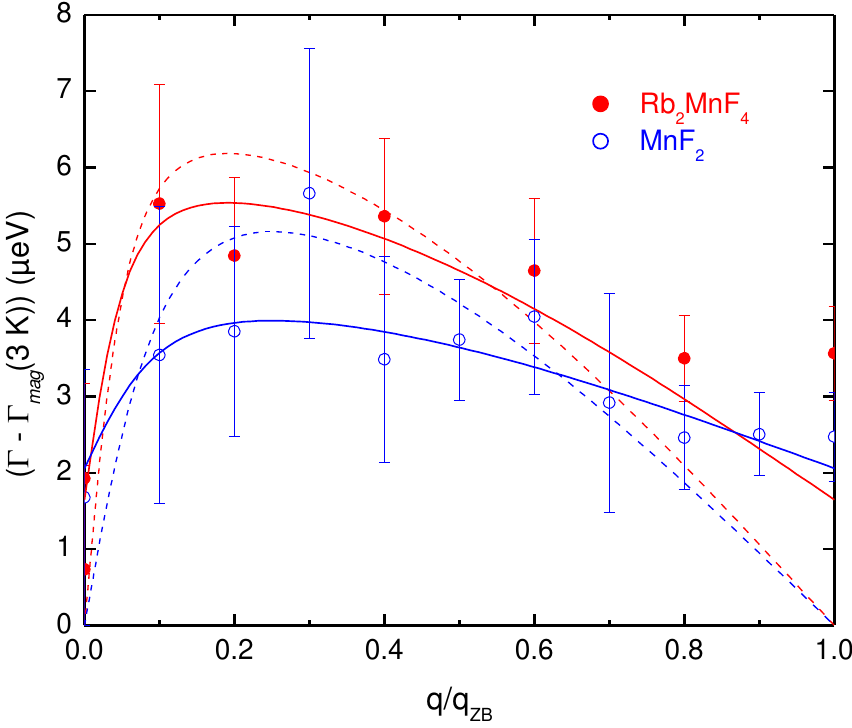}
\caption{\label{3} (Color online) Linewidths at 3 K for Rb$_{2}$MnF$_{4}$ (red) and MnF$_{2}$ (blue), with the calculated 4-magnon scattering subtracted off (this is less than 0.4 $\mu$eV for Rb$_{2}$MnF$_{4}$ and 0.05 $\mu$eV for MnF$_{2}$, for all $\mathit{q}$).  The dashed curves show fits to 2-magnon scattering (see text); the fits shown by the solid curves also allow for a constant offset.}
\end{figure}

An additional potential scattering process consists of a non-momentum-conserving collision of a magnon with a boundary (e.g. a grain boundary, antiferromagnetic domain wall, or microcrack in the crystal).  A rough estimate of the associated magnon lifetime for such a process is $\tau_{\mathit{b}} = \mathit{L}/(2\mathit{v(q)})$, where $\mathit{L}$ is the size of the grain, domain, or microcrystallite and $\mathit{v(q)}$ is the magnon velocity\cite{callaway}.  The corresponding additive contribution $\Gamma_{\mathrm{b}} = \mathit{h}\tau_{\mathit{b}}^{-1}$ to the overall linewidth exhibits a peak at low $\mathit{q}$ that is similar to that in the 3 K data (see Fig. 3).  We fitted these data to the corresponding expression after first subtracting off the calculated $\Gamma_{\mathrm{mag}}$.  Using the measured magnon dispersion data, we obtained an average $\mathit{L}$ of $0.58 \pm 0.04$ $\mu$m for Rb$_{2}$MnF$_{4}$ and $0.50 \pm 0.04$ $\mu$m for MnF$_{2}$.  The latter is consistent with the antiferromagnetic domain size of $0.63 \pm 0.05$ $\mu$m obtained from independent high-resolution neutron Larmor diffraction measurements performed on the same MnF$_{2}$ crystal at 5 K.

Under the assumption that $\mathit{L}$ remains constant at all temperatures, we calculated $\Gamma_{\mathrm{b}}$ for higher temperatures (again using the measured magnon dispersion data).  The corresponding temperature dependence is small: less than 10\% of the 3 K values, over all $\mathit{q}$, at the highest temperatures measured for MnF$_{2}$, and less than 5\% for Rb$_{2}$MnF$_{4}$.  

The fits to the 3 K data can be improved considerably by adding a small, temperature-independent constant (or such a constant plus a positive, temperature-independent slope) to the above contribution from 2-magnon scattering (see Fig. 3).  This constant is found to be $2.1 \pm 0.4$ $\mu$eV for MnF$_{2}$ and $1.7 \pm 0.3$ $\mu$eV for Rb$_{2}$MnF$_{4}$.  The corresponding values of $\mathit{L}$ are $1.4 \pm 0.8$ $\mu$m for MnF$_{2}$ and $1.0 \pm 0.2$ $\mu$m for Rb$_{2}$MnF$_{4}$.  Such an offset could reflect either an additional source of linewidth broadening or a systematic experimental error of as-yet undetermined origin.

One mechanism that could cause a line broadening of this order of magnitude is scattering from the finite width of the crystal boundaries discussed above.  The fractional effect of such scattering on the magnon energy $\mathit{E}_{\mathit{q}}$ is on the order of $(\delta\mathit{L}/\mathit{L})\mathit{E}_{\mathit{q}}$, where $\delta\mathit{L}$ is the width of the boundary.  If we take $\delta\mathit{L}$ to be one lattice constant, the corresponding change in magnon energy is on the order of 1 $\mu$eV at the zone center and 7 $\mu$eV at the zone boundary.

A mechanism that could potentially produce a linewidth that increases monotonically along the $\mathit{a}$-axis direction is 4-magnon scattering in which the electronic magnons collide with nuclear spin-waves; however, this contribution is likely to be more than an order of magnitude too small\cite{woolsey}.  Other mechanisms which may contribute, all of which are estimated to be at least an order of magnitude too small to explain the data, were summarized earlier\cite{bayrakci}.

The total calculated linewidth $\Gamma = \Gamma_{\mathrm{mag}} + \Gamma_{\mathrm{b}}$ agrees well with the data (Figs. 2-4).  In Rb$_{2}$MnF$_{4}$, the agreement worsens with increasing $\mathit{T}$ at intermediate $\mathit{q}$ (0.025 - 0.15 r.l.u.); in MnF$_{2}$, for all $\mathit{q}$, and the discrepancy is largest at intermediate $\mathit{q}$ (0.2 - 0.4 r.l.u.).  Additional scattering channels may become important at higher temperatures.  One candidate is 6-magnon (three-in/three-out) scattering, which was posited to provide the dominant contribution to the AFMR linewidth in high magnetic fields at temperatures above $\mathit{T}_{\mathit{N}}/2$ in Rb$_{2}$MnF$_{4}$ \cite{rezende1} and MnF$_{2}$ \cite{rezende2}.

Selected linewidth data from which the fitted curves to the 3 K data (Fig. 3) have been subtracted are compared in the main panel of Fig. 4b.  The exponents from (phenomenological) power-law fits for these and additional values of $\mathit{q}/\mathit{q}_{\mathit{ZB}}$ are shown in the inset.  The $\mathit{q}$-dependence of the exponents is non-monotonic; the values for the two compounds agree within error at low $\mathit{q}$, and diverge near the respective zone boundaries.  The exponent for Rb$_{2}$MnF$_{4}$ there is $2.6 \pm 0.1$, which is not too different from Kopietz' analytical result of 3 for the isotropic case (derived from approximations made to simplify the general expression of TH at the zone boundary)\cite{kopietz}, though the region of validity of his approximations lies lower in temperature than our data (see Fig. 1d).

\begin{figure}
\includegraphics[keepaspectratio=true,clip=true,trim=41mm 64mm 60mm 40mm,width=86mm]{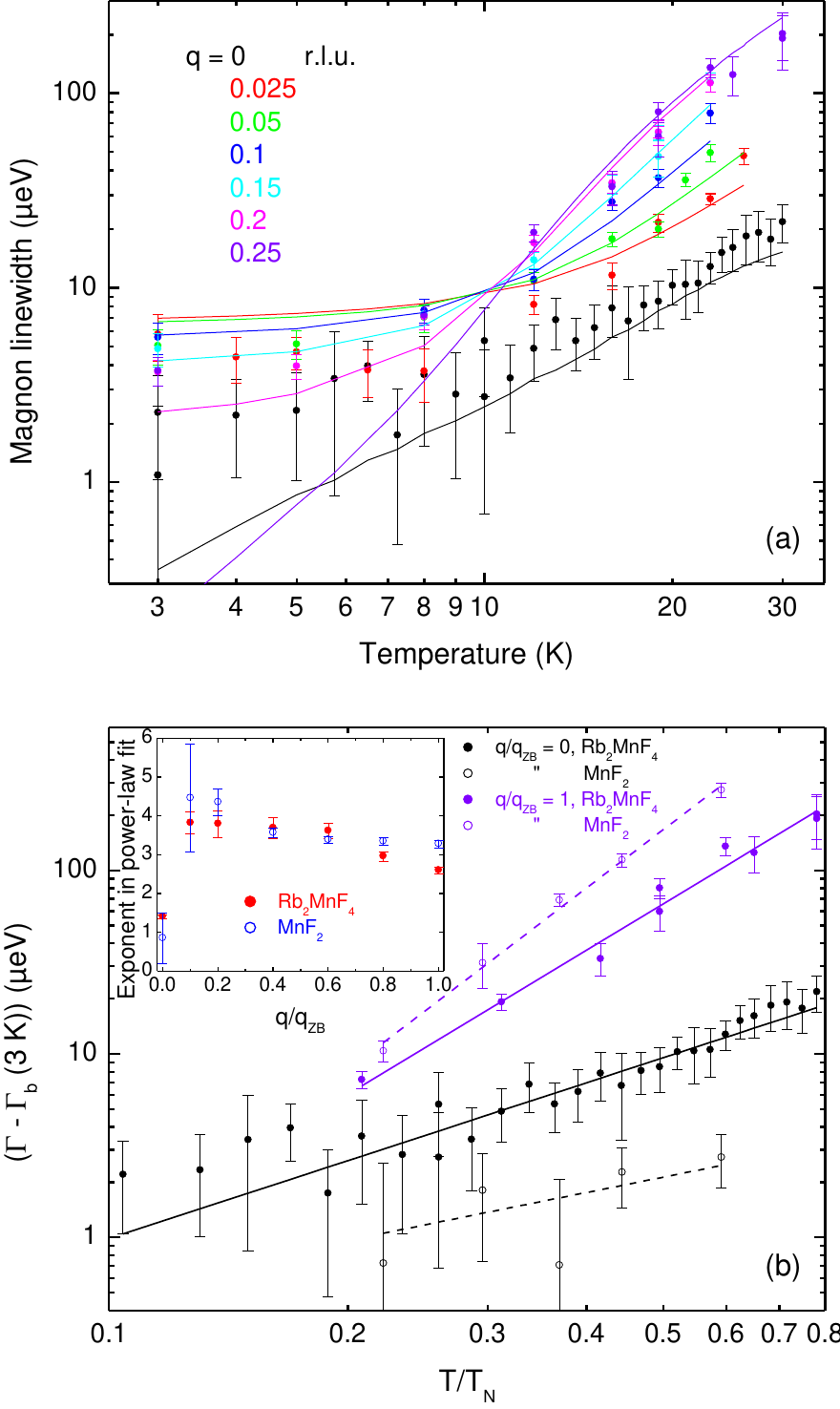}
\caption{\label{4} (Color online) (a) Linewidth data vs. $\mathit{T}$ in Rb$_{2}$MnF$_{4}$ (symbols), measured by NRSE-TAS at wavevectors along the $(\mathit{qq}0)$ direction.  The curves are calculated as described in Fig. 2.  (b) Temperature-dependent component of the magnon linewidths in Rb$_{2}$MnF$_{4}$ (open symbols) and MnF$_{2}$ (filled symbols).  The fitted contribution from 2-magnon scattering from crystal blocks at 3 K (dashed curves from Fig. 3) has been subtracted off.  The lines (solid for Rb$_{2}$MnF$_{4}$ and dashed for MnF$_{2}$) are power-law fits to the data.  The data and curves are color-coded by $\mathit{q}/\mathit{q}_{\mathit{ZB}}$.  Inset: the exponent corresponding to such power-law fits in Rb$_{2}$MnF$_{4}$ (red symbols) and MnF$_{2}$ (blue symbols).}
\end{figure}

Broadening resulting from longitudinal spin fluctuations\cite{cottam} was earlier reported to describe the magnon linewidths in MnF$_{2}$\cite{stinchcombe}.  At the temperatures considered\cite{cottam}, the longitudinal fluctuations are in fact composed to a large extent of mobile spin-waves, and thus collisions with these fluctuations are equivalent to the magnon-magnon scattering events calculated in more detail here.  The self-consistent numerical evaluations of scattering within the spin-wave gas performed in the current work thus replace both prior comparisons with the theory in 3D. 

Comparison of the hydrodynamic prediction  with our data can be done where sufficient data for a power-law fit falls within the low-$\mathit{q}$, low-$\mathit{T}$ domain of validity of the hydrodynamic theory\cite{halperin1}.  In our measurements, this applies only to MnF$_{2}$ at 25 K (the upper limit of the range of validity), where we obtain an exponent of $2.2 \pm 0.1$.  Over the range from 15 - 40 K, this exponent ranges from 1.9 - 2.2.  Hydrodynamics is expected to break down in 2D \cite{tyc}; in Rb$_{2}$MnF$_{4}$, the exponent ranges from 1.1 - 1.3 from 16 - 23 K (however, these temperatures lie above the domain of validity).

In summary, comprehensive comparison of experimental data and numerical results suggests that the processes that determine the magnon lifetime in 2D and 3D antiferromagnets are two-in/two-out magnon collisions and boundary scattering.  The difference between 2D and 3D systems is thus quantitative, not qualitative: it originates from the difference between near-cylindrical and ellipsoidal (near-spherical) surfaces of constant magnon energy with respect to the conservation of energy and momentum in 4-magnon collisions.

The seminal theories of 4-magnon scattering considered here could not previously be compared realistically and quantitatively with experimental data.  This limitation is relieved here through proper inclusion of the Umklapp scattering and the use of numerical evaluations, permitting the first complete experimental and theoretical description of magnon linewidths over broad ranges of wavevector and temperature in two and three dimensions.  Comprehensive, momentum-resolved linewidth data that are described accurately by theory enable the first calculation of the magnon-mediated thermal conductivity without adjustable parameters\cite{tobepub}.


%


\begin{acknowledgments}
We thank K. Hradil, G. Eckold, J. Major, A. Weible, I. Sorger, E. Br\"ucher, and K. Buchner for technical assistance.  This work is based upon experiments performed on the TRISP instrument operated by the MPG at the Forschungs-Neutronenquelle Heinz Maier-Leibnitz (FRM II), Garching, Germany.  The work in Stuttgart was supported by the German Science Foundation under grant SFB/TRR 80.  The work at LBNL was supported by the Director, Office of Science, Office of Basic Energy Sciences, U.S. Department of Energy under Contract No. DE-AC02-05CH11231.
\end{acknowledgments}


\begin{thebibliography}{24}
\bibitem [{HKH()}]{HKHHnote}%
  \BibitemOpen
  \href@noop {} {}\bibinfo {note} {The interaction coefficients
  $\Phi^{(1)}_{1234}$ and $\Phi^{(4)}_{1234}$ in the interaction term of the
  diagonalized Dyson-Maleev Hamiltonian used by HKHH are defined in Eqns. 2.18a
  and 2.18d of Ref. 4 and used to calculate the magnon linewidth by means of
  Eqns. 2.39 and 2.38 \cite{harris}. Our expressions for these coefficients
  differ from those of HKHH: we find $\Phi^{(1)}_{1234} =
  \gamma_{1-4}\chi_{1}\chi_{4}+\gamma_{1-3}\chi_{1}\chi_{3}+\gamma_{2-4}\chi_{2}\chi_{4}+\gamma_{2-3}\chi_{2}\chi_{3}-\gamma_{3+4-2}\chi_{2}\chi_{3}\chi_{4}-\gamma_{3+4-1}\chi_{1}\chi_{3}\chi_{4}-\gamma_{2}\chi_{2}-\gamma_{1}\chi_{1}+2\mathit{q}_{\mathit{A}}+2\mathit{q}_{\mathit{A}}\chi_{1}\chi_{2}\chi_{3}\chi_{4}$
  and $\Phi^{(4)}_{1234} =
  \gamma_{2-4}+\gamma_{1-4}\chi_{1}\chi_{2}+\gamma_{2-3}\chi_{3}\chi_{4}+\gamma_{1-3}\chi_{1}\chi_{2}\chi_{3}\chi_{4}+\gamma_{3+4-2}\chi_{3}-\gamma_{3+4-1}\chi_{1}\chi_{2}\chi_{3}-\gamma_{2}\chi_{4}-\gamma_{1}\chi_{1}\chi_{2}\chi_{4}+2\mathit{q}_{\mathit{A}}\chi_{1}\chi_{3}+2\mathit{q}_{\mathit{A}}\chi_{2}\chi_{4}$,
  where $\gamma$, $\chi$, and $\mathit{q}_{\mathit{A}}$ are defined as in Ref.
  4. Each wavevector (represented by a subscripted number) is restricted to the
  first Brillouin zone, and the sum of wavevectors is taken with no
  back-folding.}\BibitemShut {Stop}%
\end{thebibliography}
%
\providecommand{\noopsort}[1]{}\providecommand{\singleletter}[1]{#1}%

\end{document}